# Enhancing solar cell efficiency of $Al_xIn_{1-x}N$/Si heterojunctions using an a-Si buffer: a study of material, interface and device properties

M. Sun[1], R. G. Cornejo[2], M. de la Mata[3], S.I. Molina[3], B. Damilano[4], S. Valdueza-Felip[1] and F. B. Naranjo[1,*]

[1]Photonics Engineering Group. University of Alcalá, Sensors and Photonic Technologies Associated Unit, Alcalá de Henares, Spain.
[2] Instituto de Óptica Daza de Valdés, CSIC, Madrid, Spain.
[2]Departamento Ciencia de los Materiales, I.M. y Q.I., IMEYMAT, Universidad de Cádiz, Campus Río San Pedro s/n, Puerto Real, 11510 Cádiz, Spain.
[4]Université Côte d'Azur, CNRS, CRHEA, Valbonne, France.
*fernando.naranjo@uah.es

**Abstract**

This study explores the impact of an optimized amorphous silicon (a-Si) buffer layer on $Al_xIn_{1-x}$N-on-Si(100) heterojunction solar cells, with Al content varying from 0% (InN) to 55%. The buffer layer improves the structural quality of the AlInN layer, as evidenced by reduced full width at half maximum values in X-ray diffraction rocking curves around the AlInN (0002) peak. Atomic force microscopy reveals that the buffer layer does not alter surface roughness. The effectiveness of the a-Si buffer is demonstrated by an enhancement of the conversion efficiency under AM1.5G illumination from 3.3 % to 3.9 % for devices with 35 % Al. Looking at the effect of the Al content in devices with the a-Si buffer, the device with 22% Al shows the best photovoltaic performance, with a conversion efficiency of 4.1 % and a $V_{OC}$ of 0.42 V, $J_{SC}$ of 15.4 mA/cm², and FF of 63.3%. However, performance declines for Al contents above 36% due to increased resistivity and reduced carrier concentration. These findings highlight the critical role of the novel a-Si buffer layer developed by RF-sputtering and the Al content in optimizing AlInN/Si heterojunction solar cell performance.

# 1. Introduction

The pursuit for efficient, cost-effective, and durable solar cells has spurred significant research into several semiconductor materials such as a-Si, CdTe and CIGS in thin film configuration, within the so-called second-generation solar cells [1]. Further research lead to the third generation of solar cells [2], combining novel structures, like quantum dots and multijunction solar cells; and materials, such as organic materials [3] and perovskites [4]. In particular, perovskite-based solar cells in combination with ZnO and ITO show efficiency in the range of 10 % [5] and this number can be even increased up to 20 % for perovskite/Si tandem solar cells [6]. This performance lacks by the low stability of these materials [4]. On the other hand, III-Nitrides, like Indium Gallium Nitride (InGaN) and Aluminium Indium Nitride (AlInN) have emerged as a promising candidates due to their favourable electronic and optical properties, and the possibility of simple combinations with silicon. Simulations results point to efficiencies of 31 % and 23.6 %, for InGaN/Si and AlInN/Si heterojunctions, respectively [7, 8]. These numbers improve the values obtained for conventional Silicon-based solar cells [8] without using complex structures. The tunable bandgap of AlInN, ranging from the IR (0.65 eV for InN [9]) to the UV (6.05 eV for AlN [10]), depends on the Al content. By adjusting the Al composition, it is possible to align the bandgap of AlInN with the optimal absorption spectrum of solar radiation, thereby enhancing the photovoltaic conversion efficiency.

AlInN has been synthesized using several techniques, including metal-organic chemical vapor deposition (MOCVD) [11-13], molecular beam epitaxy (MBE) [14-16], elemental stacks annealing (ESA) [17,18], and reactive sputtering. Reactive sputtering can be divided into two methods: one using a combination of argon and nitrogen for deposition [19-22], and another using only nitrogen [23-27]. In our study, we employ the latter method that enhances the crystalline quality of the deposited layer, at the cost of reducing the deposition rate. The integration of AlInN with Si substrates leverages the high electron mobility and direct bandgap of AlInN while utilizing the well-established, abundant, and cost-effective silicon technology. However, the Al content in the AlInN layer influences critically the performance of these heterojunction solar cells [28]. Following this approach, Liu *et al.* reported first solar cells based on n-$Al_xIn_{1-x}N$ (x = 0.27), with a bandgap energy of 2.1 eV, on p-Si(100) heterojunctions deposited by RF sputtering, with a conversion efficiency of 1.1 % under AM-1.5G illumination [29].

At the same time, the properties and the quality of the interface between the AlInN layer and the substrate is crucial for the functionality of vertical electronic devices. However, the inherent lattice mismatch between (Al)InN and Si (-35 % for InN and -43 % for AlN) presents significant challenges, such as high defect density in the nitride layer due to the generated residual stress. This drawback can be overcame by the introduction of a well-controlled buffer layer. Several types of buffer layers have been investigated in case of nitrides for the growth on silicon substrates, such as AlN [24,25, 30], and GaN [31]. In this study, we propose the use of a thin layer of amorphous silicon (a-Si) as buffer to mitigate the lattice mismatch and minimize the defect states at the interface, which is crucial for reducing the recombination losses. The satisfactory effect of introducing an a-Si buffer on the III-nitride/Si interface has been already demonstrated in our previous work on InN nanowire solar cells [32].

The main strategies to improve the efficiency in solar cells are: optimizing the thickness, bandgap energy and bandgap alignment of the absorbers, reducing the loses by surface and interface passivation [33], reducing the contact resistivity and introducing an antireflection coating [34]. In this wok, we address the optimization strategies related to the absorber bandgap and overall quality of the n-side of the heterojunction and the interface loses reduction between the p-Si(100) substrate and the n-nitride layer. Understanding the impact of the Al content combined with the use of an a-Si buffer layer is pivotal for optimizing the design and performance of next-generation novel AlInN-on-Si solar cells.

## 2. Experimental details

a-Si and AlInN layers were deposited using a reactive (DC and RF, respectively) magnetron sputtering system (AJA International, ATC ORION-3-HV) on sapphire substrates and on *p*-doped 375-µm-thick Si(100) with resistivity of 1-10 Ω·cm . This system is equipped with 2-inch confocal magnetron cathodes of *p*-type Boron doped pure Si (5N), In (4N5) and Al (5N) targets. The background pressure of the system was in the order of $10^{-5}$ Pa and the distance between target and substrate was fixed at 10.5 cm. Substrates were chemically cleaned in organic solvents (10 minutes in acetone at 90 ºC and 10 minutes in methanol at room temperature), dried with $N_2$ and outgassed inside the chamber for 30 minutes at 550 ºC. Before deposition, the surface of the targets and the substrates were cleaned using a soft plasma etching with Ar (2 sccm and 20 W).

The a-Si buffer layer was deposited following the method detailed in our previous work [35], utilizing a pure Ar plasma with a DC power applied to the Si target of 30 W and a substrate temperature of 550 ºC. Based on previous results [32], the nominal thickness of the buffer was fixed to 15 nm. Subsequently, the $Al_xIn_{1-x}N$ layer with a nominal thickness of 260 nm was deposited at the same temperature using a pure $N_2$ plasma with a flow rate of 14 sccm and a working pressure of 0.47 Pa. The RF power applied to the Al target ($P_{Al}$) was wet to 0 (InN), 100, 125, 150, 175, 200 and 225 W, while the RF power to the In target remained constant at 30 W.

The crystalline orientation and mosaicity of the films were evaluated by high-resolution X-ray diffraction (HRXRD) measurements using a PANalytical X'Pert Pro MRD system. The thickness and morphology of the layers were studied by field-emission scanning electron microscopy (FESEM) and atomic force microscopy (AFM), respectively. A deeper study of the microstructure of the interface was carried out with transmission electron microscopy (TEM) and scanning transmission electron microscopy-energy dispersive X-ray spectroscopy (STEM-EDX). Meanwhile, Hall-Effect and optical transmission measurements were carried out at room temperature on samples co-deposited on sapphire to study the electrical and optical properties of their films. Samples were processed into devices of ~0.5 cm$^2$ area with a bottom *p*-contact on the back of the Si substrate constituted by 120 nm of Al deposited by DC sputtering and annealed at 450 ºC for 3 minutes under nitrogen atmosphere to form an ohmic contact [36]. Otherwise, the top *n*-contact consisted of 120 nm of Al deposited by RF sputtering at 225 W, except for the InN sample that required an Al contact deposited by DC sputtering to prevent adhesion and oxidation problems on the surface, due to the InN nanocolumnar morphology. Both contacts were deposited at room temperature. The contact resistivity was measured through TLM measurements, and values of $1.4 \times 10^{-3}$, $5.2 \times 10^{-1}$, $4.0 \times 10^{-2}$, 3.5, $5.2 \times 10^2$, $4.7 \times 10^3$, $2.2 \times 10^7$ $\Omega \cdot cm^2$ were obtained for samples under study (0 to 225 W of $P_{Al}$, respectively). The bottom contact resistivity is of ~4 $\Omega \cdot cm^2$. Devices were characterized by current density-voltage (J-V) curves using a 2-point probe station coupled with a source meter unit in dark and under 1 sun (1 kW/m$^2$) of standard illumination (solar simulator VeraSol class AAA) with AM1.5G spectrum.

## 3. Results and discussion

### 3.1 Structural quality

The structural quality of the $Al_xIn_{1-x}N$ layers was investigated by XRD measurements. The diffraction peaks were analysed to determine the crystal quality, Al content, and lattice parameters of the deposited layers. The samples present a wurtzite structure with the c-axis oriented along the (001) direction of the silicon substrate as shown in Fig. 1(a), showing only a single peak coming from the (0002) (Al)InN diffraction peak at 2θ = 31.3° (InN), that shifts to higher diffraction angles with the Al power. The c-lattice parameter of the nitride layer is obtained from the Bragg diffraction Law following the equation:

$$2c \cdot \sin(\theta) = n\lambda \quad (1)$$

, where c is the lattice parameter of the layer, θ is the diffraction angle, n is the diffraction order of the peak (2 in case of nitrides), and λ is the wavelength of the X-Ray radiation, being in this case, the $K_{\alpha 1}$ line of the copper, 1.54056 Å. The c-lattice parameter of the layers decreases from the theoretical value of InN (≈5.703 Å) to values closer to the theoretical value of AlN (≈4.980 Å) [37], namely 5.290 Å for an AlInN layer with an Al power of 225 W.

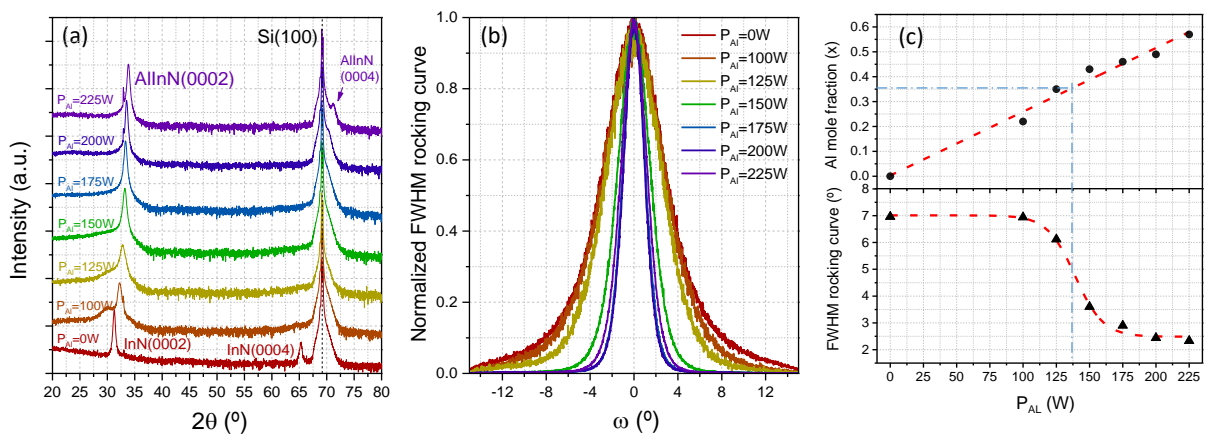

**Figure 1.** $Al_xIn_{1-x}N$ on Si(100) structures with a-Si buffer as a function of the Al power: (a) 2θ scans, (b) rocking curve of the AlInN (0002) diffraction peak and (c) calculated Al mole fraction (top) and FWHM of the rocking curve (bottom).

Assuming fully relaxed layers, we calculate the Al mole fraction (x) of the $Al_xIn_{1-x}N$ layers using the Vegard's Law, given by [38]:

$$c_{Al_xIn_{1-x}N} = xc_{AlN} + (1-x)c_{InN} \quad (2)$$

, where $c_{Al_xIn_{1-x}N}$ is the actual c value of the layer obtained from the measurement, and $c_{AlN}$ and $c_{InN}$ are the theoretical values of the lattice parameters of AlN and InN, respectively, given previously. The obtained Al mole fraction (displayed in Fig 1(c)) increases linearly with the Al power from x = 0 (InN) up to x = 0.57 ($Al_{0.57}In_{0.43}N$). These Al contents are very similar to the ones obtained in previous works for similar grown samples without a-Si buffer [26], pointing out that the introduction of the a-Si buffer layer does seem to affect the incorporation of Al in the nitride layer.

A further analysis of the rocking curve around the AlInN (0002) diffraction peak (shown in Fig. 1(b)) reveals a decrease of its full width at half maximum (FWHM) from ~7° to ~2° when increasing the Al content of the layer, which indicates an improvement of the crystalline quality of the AlInN layers with the Al incorporation. This trend has been observed in previous works of our group [26] and can be attributed to the higher energy of Al atoms ejected from the target, which increases metal adatom mobility on the substrate and improves its crystal quality, as also observed by Chen et al. [22] in AlInN samples grown by reactive sputtering. Concerning the structural quality of the samples in terms of the analyzed rocking curve, we can distinguish two regions: the low Al-content regime, for Al contents up to 35 %, and the high Al-content regime, for Al contents above this value (see Figure 2(c)).

The Full Width at Half Maximum (FWHM) of the rocking curve around the AlInN (0002) diffraction peak obtained by HRXRD measurements, in samples with a-Si buffer, is consistently lower (20-30 %) than the corresponding one in samples without buffer [36], indicating an improvement of the crystalline quality. This can be attributed to a reduction of the effect of the lattice mismatch between substrate and layer by the introduction of an amorphous layer between them.

### 3.2 Morphological and interface characterization

The morphological and surface properties of the $Al_xIn_{1-x}N$ layers and the interface between the nitride and the substrate were studied by FESEM, AFM and TEM techniques, respectively. Figure 2 (a-c) shows the SEM and AFM images of some samples under study, namely with x = 0, 0.35 (low Al-content regime) and 0.49 (high Al-content regime). It is worth saying that InN material exhibits a nanocolumnar morphology, similarly to other InN structures observed in previous works [23,32]. However, as the Al content increases, the layer morphology evolves to a compact-like one.

This change in morphology can be quantified by the root mean square (RMS) surface roughness, extracted from the 5×5 μm² AFM images, shown in Fig. 2(d), which decreases significantly from ~19 nm for InN up to 1.4 nm for Al-rich $Al_xIn_{1-x}N$ layers. This reduction in surface roughness with increasing Al content was also previously observed in our studies and is attributed to the incorporation of Al into the layers. It is known that the effective III/V flux ratio and growth temperature are crucial parameters that control the layer structure [39, 40]. Nitrogen-rich conditions (low III/V ratio) result in a nanocolumnar growth (InN), while metal-rich growth conditions (high III/V ratios) lead to the formation of a compact layers (AlInN). This transformation highlights the significant impact of Al incorporation on the structural and morphological properties of $Al_xIn_{1-x}N$ material.

The RMS surface roughness values of samples without buffer layer were similar to those of samples with buffer, as reported in [26]. This similarity suggests that the buffer layer primarily influences the crystalline quality but not affecting the surface morphology of the layers. Again, the low and high Al-content regime can be identified in Fig. 2(d), accordingly with the observed in XRD characterization (Fig. 1(c)).

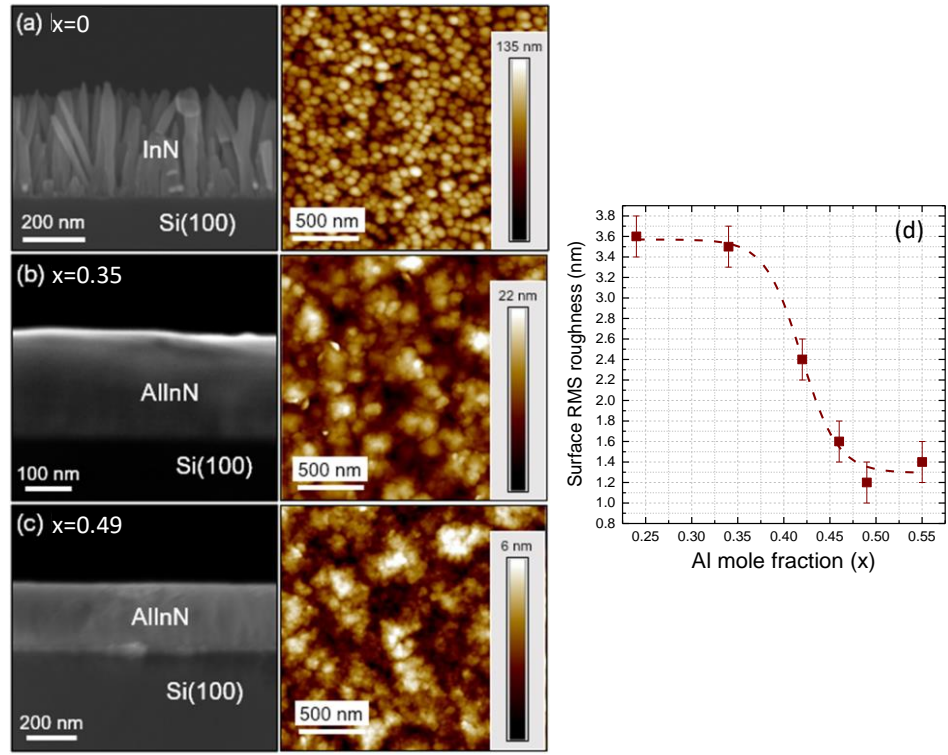

**Figure 2.** FESEM and AFM images of AlInN on Si(100) with an a-Si buffer layer for different Al contents, namely x = 0 (a), x = 0.34 (b) and x = 0.49 (c). The thickness of the samples is 360±5 nm (a), 235±5 nm (b) and 270±5 nm (b). (d) Evolution of the RMS surface roughness as a function of the Al mole fraction.

TEM analysis provides detailed insights into the microstructure and the AlInN/silicon interface. Figure 3 presents the TEM images of an $Al_xIn_{1-x}N$ sample with x = 0.32 grown directly on silicon (a) and another with a 15-nm thick a-Si buffer layer (b), along with their indexed Fast Fourier Transform (FFT). In both samples, a thin layer of ≈3 nm of $SiO_x$ is natively formed at the interface between the deposited layer (AlInN or a-Si) and the Si substrate. At the same time, the chemical composition of the samples was analyzed by STEM-EDX, showing the presence of a thin layer of ≈2 nm of $InO_x/AlO_x$ in the sample without buffer layer (not shown). This oxidized layer could affect the subsequent deposition of the AlInN layer, increasing its mosaicity and reducing the alignment of its crystal orientation with the substrate in comparison with the AlInN layer deposited using the a-Si buffer. The enhanced alignment observed when introducing the a-Si could be crucial for improving the electronic properties and overall performance of the heterojunction devices.

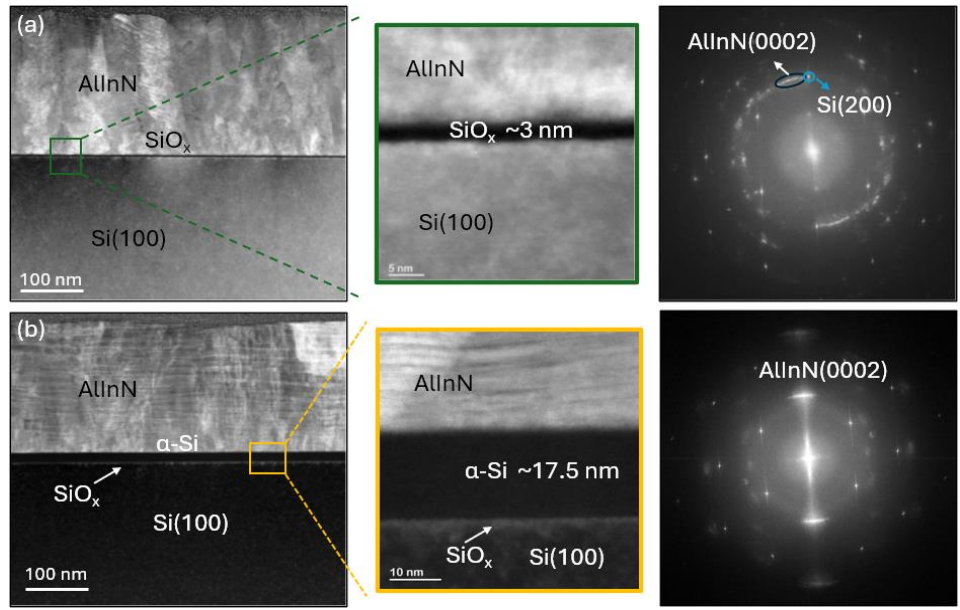

**Figure 3.** TEM and indexed FFT images of $Al_xIn_{1-x}N$ on Si(100) (a) directly grown and (b) with a-Si buffer layer with x = 0.32.

## 3.2 Optical and electrical properties

The optical properties of the $Al_xIn_{1-x}N$ samples were analysed through transmission measurements at room temperature. The inset of Fig. 4(a) shows the transmission spectra of all samples under study. As expected, the optical absorption band-edge blue shifts towards shorter wavelengths when increasing the Al content. The detailed method to estimate the bandgap energy from transmittance measurements is described in Ref. [23]. The apparent optical bandgap energy, determined by the linear fit of the squared absorption coefficient, blue shifts from 1.92 eV for the InN to 2.64 eV for $Al_{0.55}In_{0.45}N$, (Fig. 4(a)). Interestingly, the bandgap energy values of samples with the buffer layer are consistently higher than the ones of samples without the buffer layer (1.74 eV for InN – 2.57 eV for $Al_{0.56}In_{0.44}N$), even with similar Al content [36].

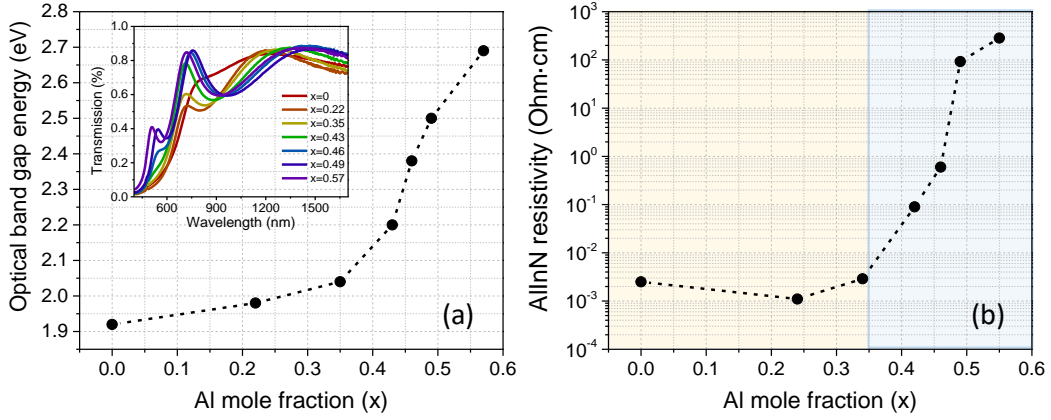

**Figure 4.** Evolution of the (a) bandgap energy and (b) resistivity of the $Al_xIn_{1-x}N$ on sapphire samples as a function of the Al mole fraction with a-Si buffer. The low and high Al-content regimes are distinguished. Inset of (a): Transmission spectra of samples under study.

The electrical properties of the samples were studied using Hall-Effect measurements at room temperature. The low and high Al-content regimes can be distinguished, with resistivity below and above $3\times10^{-3}$ Ω.cm, respectively, as shown in Fig. 4(b). These values are in good agreement with samples with similar Al content deposited under the same conditions but without the a-Si buffer [36]. The carrier concentration can be measured only for samples with low Al-content, pointing out values of $3\times10^{20}$ cm$^{-3}$ and $2\times10^{20}$ cm$^{-3}$ for samples with x = 0.24 and 0.35 of Al mole fraction, respectively. However, for high Al-content samples, the measurements became unreliable due to the high layer resistivity. A similar trend is observed in the mobility of the carriers, which is estimated to 19 cm$^{-2}$/V·s and 12 cm$^{-2}$/V·s for x = 0.24 and x = 0.35 Al-content samples, respectively. This behaviour can be attributed to the increase in bandgap energy with higher Al content, affecting both the carrier concentration and the work function of the layer, as noted by Blasco *et al.* [26]. The high carrier concentration in the $Al_xIn_{1-x}N$ layers is linked to unintentional doping from impurities such as hydrogen and oxygen during growth and the formation of defects and dislocations, consistent with findings reported in the literature from other studies [23,41,42].

The optical bandgap estimated in the samples is related to their electrical properties. Thus, the apparent bandgap energy is highly blue-shifted by the Burstein-Moss effect for the InN sample, with an apparent optical band-gap of 1.92 eV compared to the expected one for InN with very low carrier concentration (0,65 eV) [9]. The first introduction and increase of Al content, and thus the increase of the inherent bandgap energy of the material, is compensated by the reduction of the Burstein-Moss effect with the increase of the Al content due to the reduction of the carrier concentration, leading to a reduced dependence of the bandgap energy with the Al content. In this sense, a change of $\Delta E_g$ of only 250 meV for $\Delta x \approx 0.42$ (range of 0 to 0.42 Al content) is obtained. Further increase of the Al content above $x \approx 0.42$, leads to a much higher increase of $\Delta E_g$ of 370 meV for a change in Al content of only $\Delta x \approx 0.13$, being attributed a reduced effect of the Burstein-Moss effect, related with a lower carrier concentration and increased resistivity showed by the samples. This effect has been observed by other authors [43]. The obtained bandgap of 2.64 eV for the sample with $x = 0.55$ is close to the expected one for AlInN with low carrier concentration, of 2.5 eV showing that the Burstein-Moss blue-shift is reduced [44].

### 3.3 Electrical and photovoltaic device performance

A schematic representation of the developed devices under study, including the top and bottom contacts is displayed in the inset of Fig. 5(a). For each sample, four devices were fabricated and electrically characterized. However, in the paper we just show the results of the best devices for each sample. The difference between devices is related to the reproducibility of the contact metallization due to the use of a physical mask for the contacts. Figure 5 shows the current density-voltage (J-V) measurements in the dark (a) and under 1 sun of AM1.5G illumination (b) for the developed AlInN-on-Si devices with an a-Si buffer layer, as a function of the Al content.

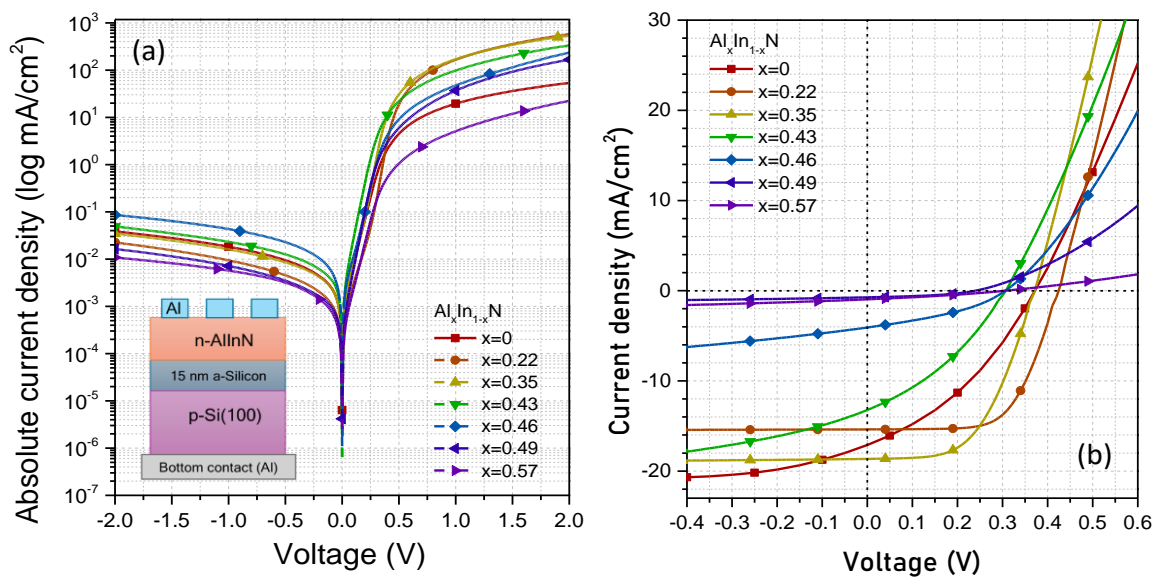

**Figure 5.** (a) J-V curves in dark of AlInN on Si(100) devices with a-Si buffer as a function of the Al content. Inset: Schematic representation of the AlInN devices. (b) J-V curves under illumination of AlInN on Si(100) devices with a-Si buffer as a function of the Al content.

The analysis of the dark J-V curves was performed using the one-diode model of the solar cell, with the equation:

$$I = I_0 \left( e^{\frac{q(V-IR_S)}{nk_BT}} - 1 \right) + \frac{V+IR_S}{R_{sh}} - I_{ph} \tag{3}$$

, where $I_{ph}$ is the photo-generated current, V is the voltage, $R_S$ is the series resistance, $R_{Sh}$ is the shunt resistance, $I_0$ is the reverse saturation current of the device, n is the ideality factor, $k_B$ is the Boltzmann constant and T is the temperature in Kelvin. The extracted parameters from the fitting to the experimental values are summarized in Table 1. For comparison, the best developed AlInN/Si(100) device without a-Si buffer layer [45] has been added to Table 1, marked with an asterisk. From these results, the first observation is that the InN device (x = 0) exhibits a correct performance with low saturation current for a nanocolumnar device, even though has a relatively high series resistance (23.2 Ω·cm²) and low shunt resistance (0.7 kΩ·cm²). For the other devices, an increase in series resistance with the Al content is observed, ranging from 2.6 Ω·cm² to 73.7 Ω·cm². This trend is consistent with the increase of the AlInN layer resistivity, as previously observed by Hall-effect measurements. On the other hand, devices exhibit high shunt resistance, in the range of 23 kΩ·cm² to 188 kΩ·cm², and low saturation currents, in the order of $10^{-6}$ to $10^{-8}$ A/cm². These values indicate a good material-substrate interface, contributing to the overall performance of the devices.

If we compare the results of the device with 35 % Al, with the best AlInN/Si(100) device developed without a-Si buffer layer (marked with an asterisk in Table 1), we observe that the fact of introducing the buffer results in superior device properties. Particularly, it leads to a reduction of the Rs, an increase of Rsh, and a reduction of the reverse leakage current, $J_0$ and the ideality factor of the junction. All these tendencies enhance the quality of the pn junction, thanks to the improved interface of the III-nitride with the Si(100) substrate.

**Table 1.** Summary of the parameters extracted from the J-V curves in dark and under 1 sun of standard illumination of the AlInN on Si(100) devices with a-Si buffer layer. Sample marked with "*" corresponds to an AlInN on Si(1000) device without a-Si buffer [25].

| $P_{Al}$ [W] | Al content [x] | Rs [Ω.cm²] | Rsh [kΩ.cm²] | $J_0$ [A/cm²] | n | Voc [V] | Jsc [mA/cm²] | FF [%] | Eff. [%] |
|---|---|---|---|---|---|---|---|---|---|
| 0 | 0 | 23.2 | 0.7 | 5.9×10⁻⁶ | 1.8 | 0.37 | 17.2 | 35.5 | 2.3 |
| 100 | 0.22 | 2.6 | 97.6 | 2.0×10⁻⁷ | 2.0 | 0.42 | 15.4 | 63.3 | 4.1 |
| 125 | 0.35 | 2.9 | 58.8 | 9.0×10⁻⁷ | 2.0 | 0.37 | 18.6 | 54.3 | 3.9 |
| *125** | *0.32* | *3.6* | *1.1* | *8.7x10⁻⁶* | *2.6* | *0.41* | *19.3* | *41.5* | *3.3* |
| 150 | 0.43 | 5.7 | 43.1 | 2.0×10⁻⁸ | 1.5 | 0.33 | 14.6 | 34.4 | 1.6 |

| 175 | 0.46 | 7.4 | 23.6 | $1.7\times10^{-7}$ | 2.0 | 0.31 | 4.0 | 36.8 | 0.5 |
| 200 | 0.49 | 14.1 | 136.6 | $7.0\times10^{-8}$ | 1.6 | 0.24 | 0.7 | 42.3 | 0.07 |
| 225 | 0.57 | 73.7 | 188.1 | $1.0\times10^{-8}$ | 2.1 | 0.31 | 0.9 | 31.0 | 0.09 |

Then, devices were characterized under 1 sun standard illumination using a solar simulator. These measurements, shown in Figure 5(b), allowed us to extract the photovoltaic parameters (open-circuit voltage, Voc, short-circuit current density, Jsc, and fill factor, FF) to calculate the conversion efficiency. All these parameters are summarized in Table 1. Best values are obtained for the device with 22 % Al, achieving a $V_{OC}$ of 0.42 V, $J_{SC}$ of 15.4 mA/cm$^2$, FF of 63.3 % and power conversion efficiency of 4.1 %, being the highest recorded to date on III-nitride on Si devices [28]. These experimental results are consistent with theoretical simulations performed in similar AlInN-on-Si devices with a bandgap energy of ~2 eV. For these devices, a conversion efficiency of approximately 4 % was predicted [8], which strongly validates the potential of III-nitride materials for high-efficiency devices.

The power conversion efficiency drops slightly to 3.9 % when increasing the Al to 35 %, However, as the Al mole fraction continues to increase, the efficiency of the devices falls significantly due to the degradation mainly of the short-circuit current and the fill factor, as a result of the high AlInN layer resistivity, that increments the contact and series resistance of the device. Together with these effects, it should be pointed out that the increase of the Al content is achieved by increasing the RF power applied to the Al target. Thus, the energy of the incoming Al atoms is also increased with the Al content, which could increase its diffusion into the buffer layer, leading to an increase of the doping of this layer. This effect has been reported in nitrides growth directly on Si substrates [46,47]. In this sense, according to our simulations, the increase of this unintentional doping of the a-Si/nitride interface leads to a reduction of the Voc, Jsc and the efficiency of the solar cell [8]. This effect of interface defects has been observed also in other material systems, like $Sb_2S_3$/CdS [48].

To demonstrate the reproducibility of the presented results, Figure 6 shows the value of power conversion efficiency, in %, obtained for the four devices fabricated with each sample, as pointed out previously. As it can be observed, a clear trend is present accordingly with the previous discussion. On the one hand, the introduction of the buffer layer leads to an enhancement of the conversion efficiency and, on the other hand, the maximum values of this magnitude are obtained in the range of Al content of 0.2 to 0.35.

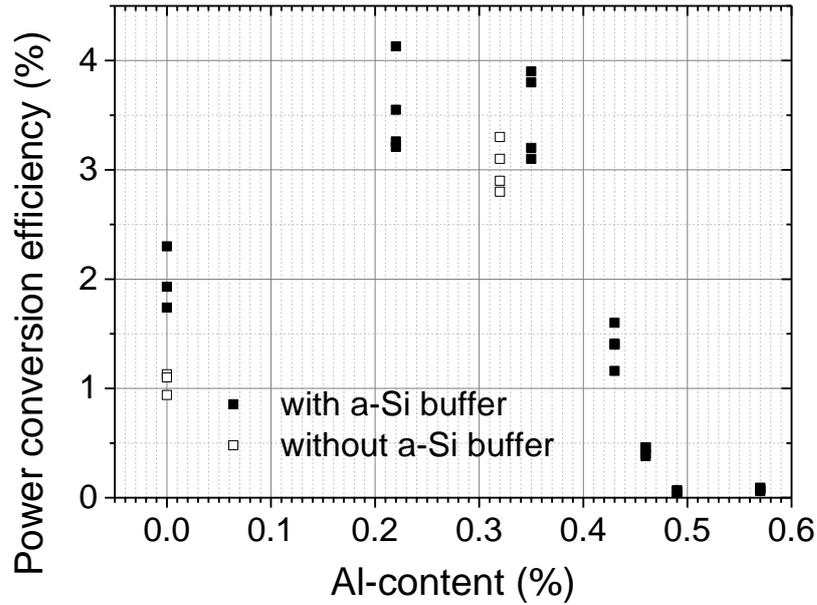

**Figure 6.** Summary of the power conversion efficiency obtained for different devices fabricated from the same sample, as explained along the text. Devices fabricated with samples developed using the a-Si buffer or without it are represented with filled and empty symbols, respectively. Data for the devices of InN (Al-content =0) without a-Si buffer were taken from samples used in study of ref. [32]

Looking to the influence of the a-Si interlayer on the photovoltaic performance of the devices, the introduction of the buffer leads to an enhancement of the fill factor of the device (see Table 1, samples deposited at 125 W) from 41.5 % to 54.3 %, and therefore of the efficiency of almost 20 % going from 3.3 % to 3.9 %. This improvement is directly related to the enhancement of the series and shunt resistances obtained when introducing the a-Si buffer, as already observed in previous studies for InN devices [32]. Figure 7 shows a simulated energy-band diagram using nextnano3 software [49] of a representative Si(100)/a-Si/(Al)InN structure with and without buffer layer (straight and dashed line, respectively). From this simulation, it can be pointed out that the a-Si buffer acts as a blocking recombination layer, reducing the recombination near the Si(100)-(Al)InN interface. This effect could be the origin of the enhancement of the device performance obtained when introducing the buffer layer, as observed by other authors [50,51].

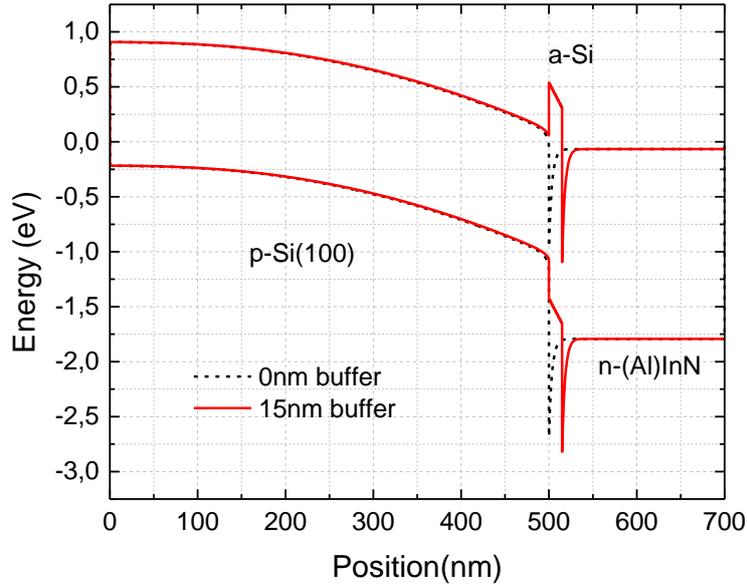

**Figure 7.** Results of the energy-band diagram simulations of the developed structures considering the nitride layer deposited directly on the p-Si(100) substrate and with the a-Si buffer layer (dash and straight lines, respectively).

## 4. Conclusions

This study systematically explores the influence of varying aluminum (Al) content and the introduction of an amorphous silicon (a-Si) buffer layer on the structural, morphological, optical, electrical, and photovoltaic properties of $Al_xIn_{1-x}N$-on-Si(100) heterojunctions. The findings highlight the significant advantages provided by the a-Si buffer layer in enhancing the crystalline quality of the AlInN layers, as evidenced by the lower FWHM values of the rocking curves compared to samples without the buffer. Adding Al transitioned the layers from nanocolumn structures to more compact films, reducing surface roughness from ~20 nm to 1.4 nm as the Al content increased. Electrically, higher Al content increased resistivity (from 2.5 mΩ·cm to 283 Ω·cm) and reduced carrier concentration. The optical analysis reveals a consistent blue shift in the bandgap energy with increasing Al content, demonstrating tunability crucial for photovoltaic applications. The structural, morphological, optical and electrical characterization of the layers allow to distinguish between a low Al-content regime (with Al content up to 35 %) and a high Al-content regime (with Al content above 35 %). The former is characterized by higher FWHM of the rocking curve of the diffraction peak around the nitride (0002) diffraction (≈7º), higher surface RMS (≈3.6 nm), and lower resistivity layers (in the range of $10^{-3}$ Ohm·cm), it being related to high carrier concentration, since these layers suffer from high Burstein-Moss effect. The contrary behavior is observed for the layers belonging to the high Al-content regime. All these properties are affecting the performance of the solar cells devices developed with these layers.

Concerning the device characterization, devices fabricated with the a-Si buffer layer showed an enhancement in power conversion efficiency, achieving a maximum efficiency of 4.1% for a device with 22% Al content. Notably, the inclusion of the buffer layer resulted in a nearly 20% improvement in efficiency for devices with 35% Al, emphasizing the critical role of the buffer in mitigating defects and enhancing interface properties. However, photovoltaic performance declined for devices with Al contents above 35%, primarily due to increased resistivity and reduced carrier concentration due to the increase of the donor activation energy expected with the increase of the bandgap energy.

Overall, the study underscores the potential of combining a-Si buffer layers with AlInN layers to optimize the performance of III-nitride-on-Si heterojunction solar cells. Theoretical simulations on n-AlInN/p-Si heterojunctions predict a maximum efficiency of 20.7 % under 1 sun of AM1.5G illumination for optimized layers with a bandgap of 2.9 eV. Moreover, such efficiency could be increased up to 25.2 % when adding a suitable antireflective coating [8]. In this sense, future work may focus on strategies such as doping the AlInN layer to reduce the contact resistance, further interface engineering to address the challenges observed at higher Al contents, the passivation of the device surfaces and the use of anti-reflecting coatings.

## Acknowledgements

Partial financial support was provided by projects SINFOTON2-CM (P2018/NMT-4326), GRISA (CM/JIN/2021-021) and CAM-project (EPU-DPTO/2020/012). This research was also funded by MCIN/AEI/10.13039/501100011033, grant number PID2022-138434OB-C53, and project PIUAH24/IA-053, from University of Alcalá.